\documentclass[preprint]{aastex61}

\received{Aug. 17, 2020}
\revised{Nov. 17, 2020}
\accepted{Nov. 30, 2020}

\submitjournal{ApJ}

\usepackage{natbib}
\usepackage{apjfonts}

\newcommand{\lae}{\mathrel{\raise .4ex\hbox{\rlap{$<$}\lower 1.2ex\hbox{$\sim$}}}}
\newcommand{\gae}{\mathrel{\raise .4ex\hbox{\rlap{$>$}\lower 1.2ex\hbox{$\sim$}}}}

\newcommand{\ixpe}{\mbox{\em IXPE\/}}

\shorttitle{Weighting of Polarizations}
\shortauthors{Marshall}

\begin{document}

\title{Multiband Weighting of X-ray Polarization Data}

\correspondingauthor{Herman L.\ Marshall}
\email{hermanm@space.mit.edu}

\author{Herman L. Marshall}
\affiliation{Kavli Institute for Astrophysics and Space Research,
 Massachusetts Institute of Technology, 77 Massachusetts Ave.,
 Cambridge, MA 02139, USA}

\begin{abstract}

An optimal estimate for Stokes parameters is derived for the
situation in X-ray astronomy where the instrument has a modulation factor that
varies significantly with energy but the signals are very weak
or mildly polarized.
For such sources, the band of analysis may be broadened in order to
obtain a significant polarization measurement.
Optimal estimators are provided for the cases of binned and unbinned data
and applied to data such as might be
obtained for faint or weakly polarized sources observed
using the Imaging X-ray Polarimetry Explorer (\ixpe).
For a sample situation, the improvement in the minimum detectable
polarization is 6-7\% using a count weighted root-mean-square of
the modulation factor, when compared to a count weighted
average.  Improving the modulation factor, such as when using a neural
network approach to IXPE event tracks, can provide additional
improvement up to 10-15\%.  The actual improvement depends on the spectral
shape and the details of the instrument response functions.

\end{abstract}


\section{Introduction}

X-ray polarimeters are still at an early stage of development
where measurements are signal limited across the entire
band of sensitivity.
Current and planned instruments
generally provide sinusoidal signals
with fractional half-amplitudes of $p\mu$ for 100$p$\%
(linear) polarized input, where $\mu$
is the modulation factor.
Often, $p$ and $\mu$ are significantly
less than unity, providing weak signals.
The minimum detectable polarization (MDP) is frequently
used to indicate how low an instrument's sensitivity might be;
when the instrument gives $p > $ MDP,
the signal should be significant at high confidence.
Most calculations of MDP select a single value for $\mu$
in order to determine the MDP even
though $\mu$ may vary significantly with
energy within the instrument's bandpass.
Here, I examine how one may account for such
variation in an optimal fashion.

X-ray polarimeters that use the photoelectric effect
such as the Imaging X-ray Polarimetry Explorer \citep[IXPE:][]{2016SPIE.9905E..17W,ixpe} or
practically any other method
including scattering \citep[e.g. X-Calibur:][]{2014JAI.....340008B,10.1117/1.JATIS.4.1.011004},
provide a signal that is probabilistically related to the
input polarization direction.
A histogram of event phase angles, $\psi$, then has a characteristic
instrument-dependent
modulation factor, $\mu$.
Ground-based calibration using sources of known polarization angle are used to
determine how $\mu$ depends on energy $E$.
Polarimeters are usually characterized by their MDP which
depends on $\mu$ and is also related to the exposure time, $T$, and
the count rates the instrument records for both the source $R_S$ and background $R_B$:

\begin{equation}
{\rm MDP}_{99} = \frac{4.292}{\mu R_S} [\frac{R_S + R_B}{T}]^{1/2}
\end{equation}

\noindent
and the level of confidence, which we will take to be 99\% \citep{weisskopf:77320E}.
When background is negligible, then

\begin{equation}
\label{eq:mdp}
{\rm MDP}_{99} = \frac{4.292}{\mu N^{1/2}}
\end{equation}

\noindent
where $N = T R_S$ is the expected number of counts in the observation.

In general, the modulation factor depends on energy.
It is defined as the ratio of the semi-amplitude of the event
histogram to the average over phase for fully polarized light.
Thus, MDP$_{99}$ can be readily computed for a small energy range but it is more difficult
to determine what value of $\mu$ to use when $\mu(E)$ is a strong function
of $E$.  One approach would be to use a count-weighted average of $\mu$:

\begin{equation}
\label{eq:mdp0}
\bar{\mu} \equiv \frac{\sum_j \mu(E_j) C_j}{N}
\end{equation}

\noindent
where $C_j$ are the counts observed in energy bin $j$ and $N = \sum_j C_j$
\cite{2012SPIE.8443E..4NE}.
Here, I compute optimal estimates of the polarization Stokes parameters
for cases where $\mu$ can vary significantly across the energy range of the detector
as long as the Stokes parameters are constant or vary slowly across the entire range.

For high signal observations of strongly polarized sources, the data can be divided into
energy bands over which the assumption of $\mu = $ constant is appropriate.
In this analysis, I will concentrate on the case when the signal is weak, so that one
requires a large bandpass in order to detect a polarization signal.
As circular polarization is not yet feasibly detectable, I will assume that Stokes $V$ is zero.
Furthermore, for simplicity, only the case where the background is negligible and
the exposure for each angular bin is the same are considered.
\citet{2015APh....68...45K} provided
an method to determine weights for the case where an instrument
must be rotated to sample the Stokes parameters fully and the sampling is not uniform.

\section{Weighting of Binned Data}

\subsection{Assumptions}

Let $i$ designate one of $n$ angular phase bins of width $\Delta \psi = 2\pi/n$ about
$\psi_i = i \Delta \psi$ ($i = 0..n-1$) and $j$ indicate the
energy bin of width $\Delta E$ about $E_j$ ($j = 1..J$).  The instrument provides counts $C_{ij}$, each
with uncertainty $\sigma_{ij}$.  I assume that the background is negligible for this analysis.
Also, for the sake of simplicity so that we may gain a little intuition about
the solution, I will assume, for now, that we have large signals in every bin so that
we may use $\chi^2$ statistics.  To make formulae a little simpler, I will set $\sigma_{ij}$ to
$\sigma$ for a few of the derivations but then generalize in \S~\ref{sec:binned}.

Since we are interested in how to combine data from different energy channels, I also assume
that the Stokes parameters $Q$ and $U$ are not functions of energy.  Otherwise, we would
not combine the channels at all.  I will not use any formalism involving
energy to channel redistribution matrix files (aka RMFs), but instead assume
that the energy response functions are narrow compared to the energy bands.  This last
point is not really restrictive because all we need in the end is a model that describes
the counts in each phase-averaged energy bin.

\subsection{The Model and Fit Statistic}

The expected number of counts in bin $i,j$ is

\begin{equation}
\lambda_{ij} = [ 1 +  \mu_j  (q \cos 2\psi_i + u \sin 2\psi_i) ] f_j A_j T \Delta E \Delta \psi
\end{equation}

\noindent
where $f_j$ is the source flux in suitable units (say ph/cm$^2$/s/keV per radian of rotation), $\mu_j$ is the modulation
factor at energy $E_j$, $A_j$ is the system effective area, $T$ is the exposure time, and $q$ and $u$
are the fractional Stokes parameters, given by
$Q = P \cos \phi_0 = q I$ and
$U = P \sin \phi_0 = u I$, respectively.  The polarization fraction
is then $p \equiv P/I = (q^2 + u^2)^{1/2}$ and the source phase angle is $\phi_0 = \tan^{-1} u/q$.
The electric vector position angle (EVPA) is $\varphi = \phi_0/2$.

We form the fit statistic from the data and the model as
\begin{equation}
\chi^2 = \sum_i \sum_j \frac{(C_{ij} - \lambda_{ij})^2}{\sigma_{ij}^2}   .
\label{eq:chi2}
\end{equation}

\noindent
To make the solution a little easier to read, I define $\alpha_j = f_j A_j T$,
$\beta_j = \alpha_j \mu_j$, $s_i = \sin 2\psi_i$, and $c_i = \cos 2\psi_i$.

At this point, I set $\sigma_{ij} = \sigma$ for all $i$ and $j$, which would apply
if the counts in each energy bin were the same and $p \ll 1$.
Eq.~\ref{eq:chi2} then becomes

\begin{equation}
\chi^2 = \frac{1}{\sigma^2} \sum_i \sum_j (C_{ij} - [ \alpha_j + q \beta_j c_i + u \beta_j s_i ] \Delta E \Delta \psi)^2   .
\label{eq:chi2b}
\end{equation}

\noindent
Because the phase angles are chosen uniformly,

\begin{eqnarray}
\sum_i s_i = \sum_i c_i = 0 \\
\label{eq:crosssums}
\sum_i s_i  c_i = 0 \\
\sum_i s_i^2 = \sum_i c_i^2 = n/2  ~~~.
\label{eq:sumsq}
\end{eqnarray}

\subsection{Solving for the Flux (Stokes I)}

Setting the derivative of $\chi^2$ with respect to $\alpha_j$ to zero yields

\begin{eqnarray}
\sum_i C_{ij} & = & n \Delta' \alpha_j + \Delta' \mu_j^2 \alpha_j \sum_i (q^2 c_i^2 + u^2 s_i^2 ) \\
	& = & n \Delta' \alpha_j + \frac{n}{2} \Delta' \mu_j^2 p^2 \alpha_j
	\label{eq:exactalpha}
\end{eqnarray}

where $\Delta' = \Delta \psi \Delta E$.
For this analysis, $\mu_j p \ll 1$, so we drop the 2nd term on the right of Eq.~\ref{eq:exactalpha}, obtaining

\begin{equation}
\label{eq:alphaj}
\hat{\alpha}_j = \frac{\sum_i C_{ij}}{2\pi \Delta E} = \frac{C_{.j}}{2\pi \Delta E}
\end{equation}

\noindent
where $C_{.j} \equiv \sum_i C_{ij}$ is the expected number of counts in channel $j$.  We also have

\begin{equation}
\label{eq:fj}
\hat{f}_j = \frac{C_{.j}}{2\pi T  A_j \Delta E}   ~~~.
\end{equation}

\noindent
Note that the units are appropriate; the factor of $1/2\pi$ comes in because the flux was
defined per radian of polarization phase space.

\subsection{Solving for the Polarization (Stokes Q and U)}

Similarly proceeding, one may derive the best estimates of $q$ and $u$:
\begin{eqnarray}
\hat{q} & = & \frac{\sum_i c_i \sum_j \beta_j C_{ij} }{\Delta' \sum_i c_i^2 \sum_j \beta_j^2} \\
\hat{u} & = & \frac{\sum_i s_i \sum_j \beta_j C_{ij} }{\Delta'\sum_i s_i^2 \sum_j \beta_j^2} ~~~~.
\end{eqnarray}

\noindent
Using the definition of $\beta_j$ and substituting Eq.~\ref{eq:alphaj}, we have
\begin{eqnarray}
\label{eq:q}
\hat{q} = \frac{n \sum_i c_i \sum_j \mu_j  C_{.j} C_{ij}}{\sum_i c_i^2 \sum_j \mu_j^2 C_{.j}^2} = 
	\frac{2 \sum_j \mu_j C_{.j} \sum_i c_i C_{ij}}{\sum_j \mu_j^2  C_{.j}^2} = \sum_i c_i x_i \\
\hat{u} = \frac{n \sum_i s_i \sum_j \mu_j C_{.j} C_{ij} }{\sum_i s_i^2 \sum_j \mu_j^2  C_{.j}^2} =
	\frac{2 \sum_j \mu_j C_{.j} \sum_i s_i C_{ij}}{\sum_j \mu_j^2  C_{.j}^2} = \sum_i s_i x_i 
\label{eq:u}
\end{eqnarray}

\noindent
where the definition of $C_{.j}$ and Eq.~\ref{eq:sumsq} are used and

\begin{equation}
x_i \equiv \frac{2 \sum_j \mu_j C_{.j} C_{ij}}{\sum_j \mu_j^2 C_{.j}^2} ~~~.
\end{equation}

\noindent
Also,

\begin{equation}
\tan(\hat{\phi_0}) = \frac{\sum_j \mu_j C_{.j} \sum_i s_i C_{ij}}{\sum_j \mu_j C_{.j} \sum_i c_i C_{ij}} = \frac{\sum_i s_i x_{i}}{\sum_i c_i x_{i}}
\end{equation}

\noindent
\begin{equation}
\hat{p} = [ ( \sum_i c_i x_i )^2 + ( \sum_i s_i x_i )^2 ]^{1/2} = | \digamma_2(x) |
\end{equation}

\noindent
where $\digamma_2(x)$ is the 2nd term of the Fourier transform of $x_i$ (i.e., the $2\theta$ modulation
portion of the phase binned counts weighted by a channel-dependent term).
Note that $p \mu_j C_{.j}$ is the maximum modulated signal in channel $j$,  and $\mu_j C_{.j}$ is the maximum modulated signal when $p = 1$.

The best estimates of $q$ and $u$ for a single
energy channel $j$ are obtained from Eqs~\ref{eq:q} and \ref{eq:u} by removing the summation over $j$, giving

\begin{eqnarray}
\label{eq:q1}
\hat{q}_j = \frac{2 \sum_i c_i C_{ij}}{\mu_j  C_{.j} } \\
\hat{u}_j = \frac{2 \sum_i s_i C_{ij}}{\mu_j  C_{.j} }
\label{eq:u1}
\end{eqnarray}

\noindent
We can check these formulae
by modeling with $C_{ij} = C_{.j}/n (1 + \eta \mu_j \cos 2\psi_i ) = C_{.j}/n (1 + \eta \mu_j c_i )$ as the response of the detector to
a signal that is $100\eta$\% polarized at an EVPA $= \varphi = \phi_0 = 0$.
Substituting the model into Eqs.~\ref{eq:q1} and \ref{eq:u1}  gives
$\hat{q}_j = \eta$ and $\hat{u}_j = 0$, as expected, using Eqs.~\ref{eq:crosssums} and \ref{eq:sumsq}.

\subsection{Bandpass Weighting}

Solving Eqs.~\ref{eq:q1} and \ref{eq:u1} for $\sum_i c_i C_{ij}$ and $\sum_i s_i C_{ij}$, respectively,
then Eqs.~\ref{eq:q} and \ref{eq:u} can be rewritten as

\begin{eqnarray}
\hat{q} = \sum_j W_j \hat{q}_j \\
\hat{u} = \sum_j W_j \hat{u}_j
\end{eqnarray}

\noindent
where

\begin{equation}
W_j = \frac{\mu_j^2 C_{.j}^2}{\sum_k \mu_k^2 C_{.k}^2}
\end{equation}

\noindent
Because $\sum_j W_j = 1$, the $W_j$ values can be identified as weights to apply to the individual
bandpass values in order to obtain optimal, global values, as long as the polarization angle
and fraction do not depend significantly on energy.  The relative weight of channel $j$ compared
to channel $k$ is just $(\mu_j/\mu_k)^2 (C_{.j}/C_{.k})^2$, the ratio of the variances of the total
counts in the two channels weighted
by the square of the relevant modulation factors.

\subsection{Stokes Parameter Uncertainties, Binned Case, Constant Uncertainties}

\label{sec:constuncert}
The diagonal terms of the Fisher matrix give the parameter uncertainties for $f_j$, $q$, and $u$, as the
off-diagonal derivatives of Eq.~\ref{eq:chi2b} are all zero or very small when $p \mu \ll 1$.
Because of the assumption that the bandpasses are arranged to have equal
variances, $\sigma^2$, $C_{.j} = n \sigma^2 = N/J$, giving 

\begin{eqnarray}
\sigma_{f_j}^2 & = & \frac{2}{\frac{\partial^2 \chi^2}{\partial f_j^2}} = 
	\frac{2}{(T A_j)^2 \frac{ \partial^2 \chi^2}{\partial^2 \alpha_j^2}} = \frac{\sigma^2}{(T A_j \Delta')^2 n}
	= \frac{C_{.j}}{(2 \pi T A_j \Delta E)^2}\\
\sigma_q^2 & = & \frac{2}{\frac{\partial^2 \chi^2}{\partial q^2}} = 
	\frac{2}{2 (\Delta')^2 \sum_i c_i^2 \sum_j \alpha_j^2 \mu_j^2} = 
	\frac{2 n \sigma^2}{ \sum_j\mu_j^2 C_{.j}^2} = \frac{2 J}{N \sum_j \mu_j^2}  \\
\sigma_u^2 & = & \frac{2}{\frac{\partial^2 \chi^2}{\partial u^2}} =  
	\frac{2}{2 (\Delta')^2 \sum_i s_i^2 \sum_j \alpha_j^2 \mu_j^2} = 
	\frac{2 n \sigma^2}{ \sum_j\mu_j^2 C_{.j}^2} = \frac{2 J}{N \sum_j \mu_j^2} ~~~~.
\end{eqnarray}

\noindent
In order to relate to the Stokes parameters' uncertainties to MDP$_{99}$, consider
the case of a single channel, where $J = 1$ and  $\mu_j = \mu$, so

\begin{equation}
\sigma_q^2 = \sigma_u^2 = \frac{2}{\mu^2 N}~~~~.
\end{equation}

\noindent
MDP$_{99}$ is then just scaled to the uncertainty in $q$ or $u$:

\begin{equation}
{\rm MDP}_{99} = \frac{4.292}{\mu N^{1/2}} = 3.035 \sigma_q = 3.035 \sigma_u ~~~~.
\end{equation}

\noindent
Due to the symmetry between $q$ and $u$ in the absence of a signal, we can determine the
MDP for uniformly weighted polarimeter counts when there are several bandpasses with
differing modulation channels:

\begin{equation}
{\rm MDP}_{99} = 3.035 \sigma_q = \frac{4.292 J}{N \sum_j \mu_j^2} = \frac{4.292}{(\bar{\mu^2} N )^{1/2}} ~~~~.
\label{eq:mdp2}
\end{equation}

\noindent
Eq.~\ref{eq:mdp2} indicates that the optimal modulation factor to use
for a multi-band data set with constant counts per energy bin
is the RMS of the set of $\mu_j$ values, not the mean.

\subsection{Using Data Uncertainties, Binned Case}
\label{sec:binned}

While commonly practiced, it is not always appropriate to set spectral bins based on
the observed counts in order to obtain a constant number of counts per spectral bin.
One reason to avoid such an approach is that the bins may end up too wide to satisfy
the requirement that the modulation factor be constant across the bin.
So, now we return to the general case where uncertainties depend on the energy bin $j$.
With statistical weighting by $1/\sigma_{ij}^2$, the weighted sums over $c_i$, $s_i$, and $c_i s_i$
are no longer identically equal to zero as the uncertainties track the counts per bin, which will correlate
with $\psi_i$ for $p \mu_j > 0$.  However, when considering the special case of weakly
modulated signals, there can be simple solutions.

The best values of $\alpha_j$, $q$, and $u$ are found by setting $\nabla \chi^2 = 0$.
Using Eq.~\ref{eq:chi2} for $\chi^2$ gives

\begin{eqnarray}
\Delta' \hat{\alpha}_j  \sum_i (1 + \hat{q} \mu_j c_i + \hat{u} \mu_j s_i)^2 /\sigma_{ij}^2
	& = &  \sum_i C_{ij} (1 + \hat{q} \mu_j c_i + \hat{u} \mu_j s_i) /\sigma_{ij}^2, ~~j=1..J\\
\sum_j \mu_j^2 \hat{\alpha}_j^2 \sum_i (\hat{q} c_i^2 + \hat{u} c_i s_i)/ \sigma_{ij}^2 & = & 
	\sum_j \mu_j \hat{\alpha}_j \sum_i (C_{ij}/\Delta' -   \hat{\alpha}_j) c_i / \sigma_{ij}^2 \\
\sum_j \mu_j^2 \hat{\alpha}_j^2 \sum_i (\hat{q} c_i s_i + \hat{u} s_i^2)/ \sigma_{ij}^2 & = & 
	\sum_j \mu_j \hat{\alpha}_j \sum_i (C_{ij}/\Delta' -   \hat{\alpha}_j) s_i / \sigma_{ij}^2
\end{eqnarray}

\noindent
These equations will not be tractable in general but we can derive approximate solutions
for $p \mu_j \ll 1$ and we may drop the $c_i s_i$ terms that are small
compared to the $c_i^2$ and $s_i^2$ terms:

\begin{equation}
\hat{\alpha}_j = \frac{1}{\Delta'} \frac{\sum_i C_{ij} /\sigma_{ij}^2}{\sum_i 1 /\sigma_{ij}^2} \equiv \frac{\bar{C}_j}{\Delta'}
\end{equation}

\begin{eqnarray}
\label{eq:q2}
\hat{q} = \frac{\sum_j \mu_j \bar{C}_j \sum_i c_i (C_{ij}-\bar{C}_j) / \sigma_{ij}^2}{\sum_j \mu_j^2 \bar{C}_j^2 \sum_i c_i^2 / \sigma_{ij}^2}  \\
\hat{u} = \frac{\sum_j \mu_j \bar{C}_j \sum_i s_i (C_{ij}-\bar{C}_j) / \sigma_{ij}^2}{\sum_j \mu_j^2 \bar{C}_j^2 \sum_i s_i^2 / \sigma_{ij}^2}
\label{eq:u2}
\end{eqnarray}

\noindent
where the quantity $\bar{C}_j$ is a variance-weighted average of the counts in channel $j$.
When $\sigma_{ij}^2 = C_{ij}$,
then $\bar{C}_j = n/(\sum_i C_{ij}^{-1} )$.
The single channel estimates of the $q$ and $u$ values are 

\begin{eqnarray}
\label{eq:qj2}
\hat{q_j} & = &  \frac{\sum_i c_i (C_{ij}-\bar{C}_j) / \sigma_{ij}^2}{\mu_j \bar{C}_j \sum_i c_i^2 / \sigma_{ij}^2}  \\
\hat{u_j} & = & \frac{\sum_i s_i (C_{ij}-\bar{C}_j) / \sigma_{ij}^2}{\mu_j \bar{C}_j \sum_i s_i^2 / \sigma_{ij}^2}  ~~~~.
\label{eq:uj2}
\end{eqnarray}

Again, we can solve Eqs.~\ref{eq:qj2} and \ref{eq:uj2} for the terms that sum over $C_{ij}/\sigma_{ij}^2$  in
the numerators to
obtain estimates of $q$ and $u$ in terms of single band estimates

\begin{eqnarray}
\label{eq:q3}
\hat{q} & = & \sum_j {w'}_j \hat{q}_j  \\
\hat{u} & = & \sum_j {w''}_j \hat{u}_j
\label{eq:u3}
\end{eqnarray}

\noindent
where

\begin{eqnarray}
{w'}_j = \frac{\mu_j^2 \bar{C}_j^2 \sum_i c_i^2/\sigma_{ij}^2}{\sum_k \mu_k^2 \bar{C}_k^2 \sum_i c_i^2/\sigma_{ik}^2} \\
{w''}_j = \frac{\mu_j^2 \bar{C}_j^2 \sum_i s_i^2/\sigma_{ij}^2}{\sum_k \mu_k^2 \bar{C}_k^2 \sum_i s_i^2/\sigma_{ik}^2}
\end{eqnarray}

\noindent
and, as before, $\sum_j {w'}_j = \sum_j {w''}_j = 1$.
These are the weights to be applied to single band results to combine them optimally.

\subsection{Stokes Parameter Uncertainties, Binned Case}

Again, uncertainties in the derived values can be estimated under the assumption $p\mu \ll 1$
so that cross-terms in the Fisher matrix are negligible.  Then, 

\begin{eqnarray}
\sigma_{f_j}^2 & = & \frac{2}{\frac{\partial^2 \chi^2}{\partial f_j^2}} = \frac{1}{(T A_j \Delta')^2 \sum_i 1/\sigma_{ij}^2} \\
\sigma_q^2 & = & \frac{2}{\frac{\partial^2 \chi^2}{\partial q^2}} = \frac{1}{\sum_j  \mu_j^2 \bar{C}_j^2 \sum_i c_i^2/ \sigma_{ij}^2} \\
\sigma_u^2 & = & \frac{2}{\frac{\partial^2 \chi^2}{\partial u^2}} = \frac{1}{\sum_j  \mu_j^2 \bar{C}_j^2 \sum_i s_i^2/ \sigma_{ij}^2}
\end{eqnarray}

\noindent
For weak modulations, the counts in channel $j$ are approximately independent of phase bin, so
we may approximate $\sigma_{ij}^2$ by $\sigma_j^2 / n$, where $\sigma_j^2 \approx C_{.j}$ is the variance
in total counts in channel $j$, $C_{.j}$.  Now $\bar{C}_j = C_{.j}/n$, giving

\begin{eqnarray}
\sigma_q^2 & = &  \frac{n}{\sum_j  \mu_j^2 C_{.j}^2/\sigma_j^2 \sum_i c_i^2} = 
	\frac{2}{\sum_j  \mu_j^2 C_{.j}^2/\sigma_j^2} \approx \frac{2}{\sum_j  \mu_j^2 C_{.j}}\\
\sigma_u^2 & = & \frac{n}{\sum_j  \mu_j^2 C_{.j}^2/\sigma_j^2 \sum_i s_i^2} = 
	\frac{2}{\sum_j  \mu_j^2 C_{.j}^2/\sigma_j^2} \approx \frac{2}{\sum_j  \mu_j^2 C_{.j}}
\end{eqnarray}

Following the computation of MDP$_{99}$ as in \S~\ref{sec:constuncert}, MDP$_{99}$ in this case is

\begin{eqnarray}
\label{eq:mdp3a}
{\rm MDP}_{99} & = & 3.035 \sigma_q = \frac{4.292}{( \sum_j  \mu_j^2 C_{.j}^2/\sigma_j^2)^{1/2}} \\
	& \approx & \frac{4.292}{(\sum_j  \mu_j^2 C_{.j})^{1/2}} ~~~~.
\label{eq:mdp3}
\end{eqnarray}

\noindent
Eq.~\ref{eq:mdp3} indicates that the modulation factor to use in Eq.~\ref{eq:mdp}
that is optimal for a multi-band data set is the count-weighted
RMS of the set of $\mu_j$ values, $\sum_j \mu_j^2 C_{.j} / N$, not the count-weighted mean
(Eq.~\ref{eq:mdp0}).  The choice to use Eq.~\ref{eq:mdp3a} or Eq.~\ref{eq:mdp3} depends on whether
the variances are dominated by the total counts in the energy band; if there is
significant non-Poisson variance, then Eq.~\ref{eq:mdp3a} is
will be somewhat more accurate.

\section{Weighting of Unbinned Data}

In general, an unbinned likelihood analysis is more appropriate
than the use of $\chi^2$ statistics for fitting X-ray event data
due to the Poisson nature of the counting statistics.
Furthermore, binning details often involves user discretion.
Though normally requiring numerical methods to determine
model parameters, the assumption of weak modulations makes it
possible to obtain good estimates.

\subsection{Likelihood Formulation}

Assume that there are $N$ events, with energies and instrument phases $(E_i, \psi_i)$.
At energy $E_i$, the modulation factor is $\mu_i$, the instrument effective area is $A_i$,
and the intrinsic source flux is $f_i = f(E_i)$ based on the spectral model of the source.
The event density in a differential energy-phase element $dE d\psi$ about $(E, \psi)$ is

\begin{equation}
\lambda(E, \psi) = [ 1 +  \mu_E  (q \cos 2\psi + u \sin 2\psi) ] f_E A_E T dE d\psi
\end{equation}

\noindent
and the log-likelihood for a Poisson probability distribution of events, $S = -2 \ln L$, is

\begin{eqnarray}
S & = & -2 \sum_i \ln \lambda(E_i, \psi_i) + 2 T \int f_E A_E dE  \int_0^{2\pi} [ 1 +  \mu(E)  (q \cos 2\psi + u \sin 2\psi) ]  d\psi \\
 & = & -2 \sum_i \ln f_i -2 \sum_i  \ln( 1 +  q \mu_i \cos 2\psi_i + u \mu_i \sin 2\psi_i) + 4 \pi T \int f_E A_E dE + {\rm constant}
\label{eq:unbin}
\end{eqnarray}

\noindent
Note that eq.~\ref{eq:unbin} has two terms
that depend only on the parameters of $f_E$ and the remaining
interesting term depends only on
the parameters of $q$ and $u$.  If there are no parameters in common, in the
case where $q$ and $u$ do not depend on the source flux and we are
not fitting (yet) for parameters of the spectral model, then the log-likelihood
for the polarization parameters is merely

\begin{equation}
S(q,u) =  -2 \sum_i  \ln ( 1 +  q \mu_i \cos 2\psi_i + u \mu_i \sin 2\psi_i)   .
\label{eq:like}
\end{equation}

\subsection{Estimating $q$ and $u$ for the Unbinned Case}

We can define $c_i = \mu_i \cos 2\psi_i$ and
$s_i = \mu_i \sin 2\psi_i$ (unlike the case for binned data).  Setting
$\partial S/\partial q = 0$ and $\partial S/\partial u = 0$
to find the best estimates of $q$ and $u$ gives 

\begin{eqnarray}
\label{eq:gen1}
0 = \sum_i \frac{c_i}{1 + \hat{q}c_i + \hat{u}  s_i} = \sum_i w_i c_i\\
0 = \sum_i \frac{s_i}{1 + \hat{q}c_i + \hat{u}  s_i} = \sum_i w_i s_i
\label{eq:gen2}
\end{eqnarray}

\noindent
where $w_i \equiv (1 + \hat{q} c_i + \hat{u}  s_i)^{-1}$.  It can be shown
that $\sum_i w_i = N$.  These two equations apply under quite general
circumstances but require numerical solution.  However, for $\hat{q} \ll 1$ and $\hat{u}  \ll 1$,
then $w_i \approx 1 - \hat{q} c_i - \hat{u} s_i$, so eqs.~\ref{eq:gen1} and \ref{eq:gen2} become

\begin{eqnarray}
\sum c_i = \hat{u} \sum c_i s_i + \hat{q} \sum c_i^2\\
\sum s_i = \hat{u} \sum s_i^2 + \hat{q} \sum c_i s_i
\end{eqnarray}

\noindent
where all sums are implicitly over $i$.  If the source is polarized, then the phase angles are not random, so the sums
over the sine or sine-cosine terms
are not identically zero.
Fortunately, the pair of linear equations can be solved, giving

\begin{eqnarray}
\hat{q} =  \frac{\sum c_i \sum s_i^2 - \sum s_i \sum c_i s_i}{\sum s_i^2 \sum c_i^2 - (\sum c_i s_i)^2}\\
\hat{u} =  \frac{\sum s_i \sum c_i^2 - \sum c_i \sum c_i s_i}{\sum s_i^2 \sum c_i^2 - (\sum c_i s_i)^2}
\end{eqnarray}

\noindent
In keeping with the assumed approximations of small $q$ and $u$,
we expect the sine-cosine terms to be small compared to the sums over $s_i^2$
or $c_i^2$, giving

\begin{eqnarray}
\hat{q} \approx \frac{\sum c_i}{ \sum c_i^2}\\
\hat{u} \approx \frac{\sum s_i}{ \sum s_i^2}
\end{eqnarray}

It should be simple to compute these sums for all the events in an observation in order to
obtain a first estimate of the sizes of the fractional Stokes parameters, $q$ and $u$.
These estimates provide some accounting for the variation of the modulation factor by
using it as a weight on each phase term. 
Furthermore, the solution indicates a simple way to estimate the dependences of $q$ and $u$
on $E$ by merely creating the sums over suitably small energy ranges so one may obtain
a non-parametric estimate of the $(q,u)$ energy dependence, assuming that the values are
small but measurable.

The RMF does not enter this analysis, making it relatively easy to examine a data set
in a preliminary fashion before attempting detailed fitting that might require an
iterative fitting method that accounts for energy redistribution.
Technically, assigning a value of $\mu$ to an event
requires knowledge of the event's true energy, which
is uncertain because of the probabilistic
mapping between the detector signal and energy.
However, as long as the modulation factor
varies slowly compared to the detector response function, this approach may still
be useful and give a valid impression as to how to devise a physical model.

\subsection{Uncertainties, Unbinned Case}

Finally, the uncertainties on $q$ and $u$ are estimated from the 2nd derivatives
of eq.~\ref{eq:like}:

\begin{eqnarray}
\frac{\partial^2 S}{\partial u^2} = 2 \sum w_i^2 s_i^2\\
\frac{\partial^2 S}{\partial q^2} = 2 \sum w_i^2 c_i^2\\
\frac{\partial^2 S}{\partial u \partial q} = 2 \sum w_i^2 c_i s_i
\end{eqnarray}

\noindent
and using $w_i \approx 1$ and $\sum w_i^2 c_i s_i \approx 0$,
one obtains simple estimates for the Stokes uncertainties:

\begin{eqnarray}
\sigma_q^2 \approx \frac{2}{\frac{\partial^2 S}{\partial q^2}} \approx \frac{1}{\sum c_i^2}\\
\sigma_u^2 \approx \frac{2}{\frac{\partial^2 S}{\partial u^2}} \approx \frac{1}{\sum s_i^2}
\end{eqnarray}

\noindent
Remember that the $c_i$ and $s_i$ values have the modulation factor included
in this situation, so the denominators
are not $N/2$.  To estimate the denominators, we note that
\begin{eqnarray}
\sum c_i^2 + \sum s_i^2 = \sum \mu_i^2 = N \bar{\mu^2}  ~~,
\end{eqnarray}
where $\bar{\mu^2}$ is the
count-weighted average of $\mu^2$.
Furthermore, $c_i$ and $s_i$ should be similarly distributed
when $p \ll 1$, giving
\begin{eqnarray}
\sum c_i^2 \approx \sum s_i^2 \approx \frac{N}{2} \bar{\mu^2}
\end{eqnarray}

\noindent
Finally, as in \S~\ref{sec:binned}, we make the connection to MDP$_{99}$:

\begin{equation}
\label{eq:mdp4}
{\rm MDP}_{99} = 3.035 \sigma_q  \approx \frac{3.035}{(\sum_i c_i^2)^{1/2}}
	 \approx \frac{4.292}{(\bar{\mu^2} N)^{1/2}} ~~~~.
\end{equation}

\noindent
As in the binned case,
the modulation factor to use in Eq.~\ref{eq:mdp}
that is optimal for unbinned data set is the count-weighted
RMS of the set of $\mu_i$ values, not the count-weighted mean
(Eq.~\ref{eq:mdp0}).

\section{Example Application and Summary}

Armed with an appropriate weighting scheme, it is useful to determine how much
the MDP may improve using it.
To determine the effect of optimal binning, the results of an observations
can be computed using assumed functions for
the spectral model, effective area, and
$\mu(E)$.  It is important to get the dependences of these functions
on energy approximately correct; the normalizations are not relevant when just
selecting a total number of events for each simulated observation.

\begin{figure}
    \centering
    \includegraphics[width=15cm]{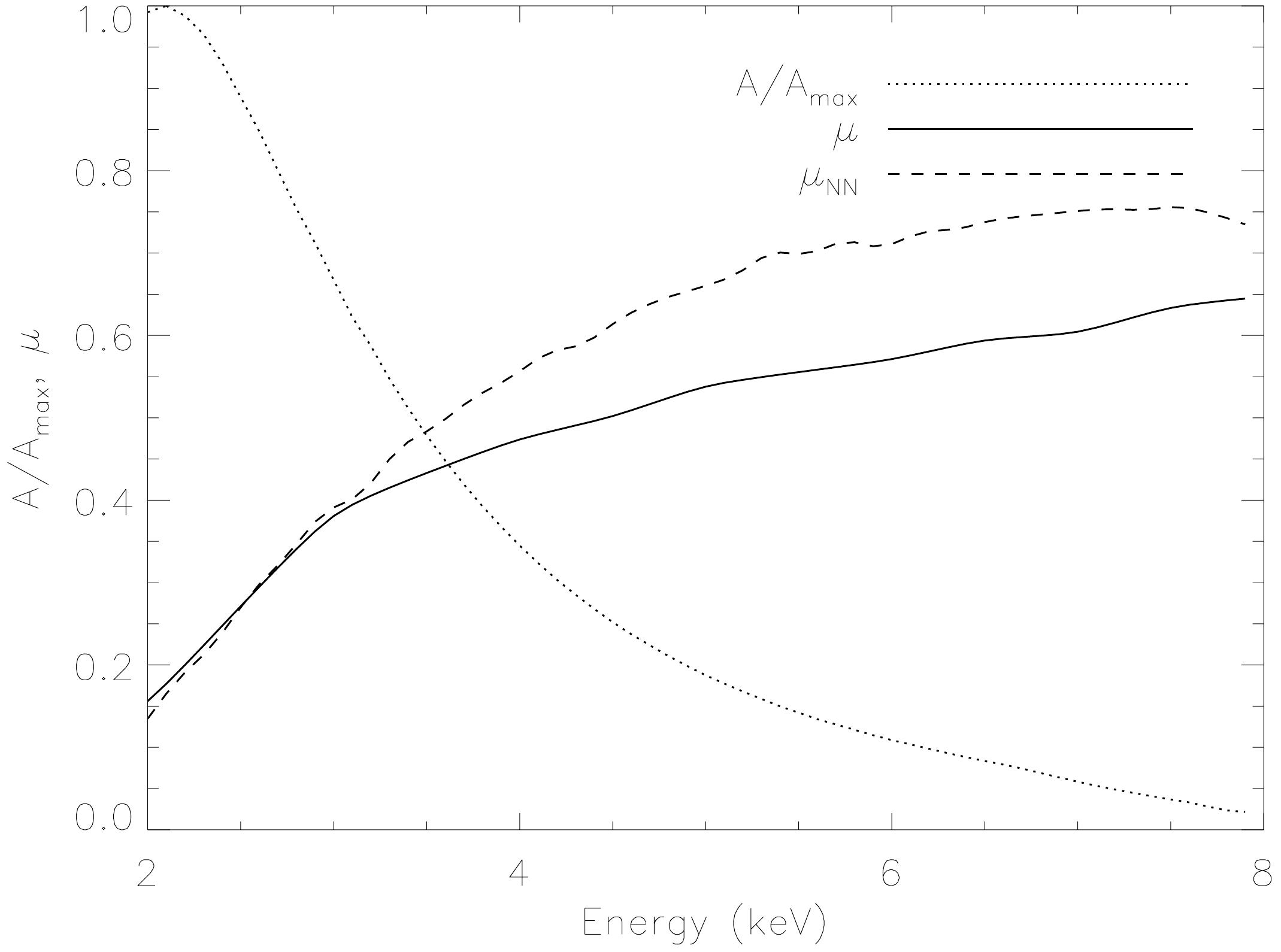}
 \caption{ 
Representative functions for the effective area, $A$, and modulation
  function, $\mu$ for an X-ray polarimeter.  The effective area curve is illustrative,
  for an instrument like IXPE \citep{2012SPIE.8443E..4NE,ixpe}.  The
  effective area is normalized arbitrarily to the maximum value.
  The solid line gives the standard $\mu(E)$ for IXPE
  (A.\ Di Marco, {\it et al.}, in prep.).  The dashed curve gives the $\mu(E)$ curve
  from a neural network method with weighting parameter $\lambda = 2$ \cite{PEIRSON2021164740}.
 }
\label{fig:functions}
\end{figure}

Fig.~\ref{fig:functions} shows assumed models of the effective area,
normalized to 1 at the maximum value, and of $\mu(E)$, both
for an instrument like IXPE \cite{ixpe}.
The standard modulation curve for IXPE uses a moment method
with ellipticity exclusion \citep{2003SPIE.4843..383B};
the data are from A.\ Di Marco, {\it et al.}, in preparation
\citep[see also][]{2016SPIE.9905E..17W}.
The effective area is approximate
\citep[as in][]{2012SPIE.8443E..4NE} -- only the shape is important for
this analysis.
For a given
total number of counts, $N$, two versions of MDP$_{99}$ were computed
using the area and modulation curves for each of two assumed spectral models.
The spectral model is a power law with spectral index $\Gamma = 0$ or 2,
where $f_E \propto E^{-\Gamma}$.
For case 1, MDP$_{99}$ is determined using a count weighted modulation
factor given by Eq.~\ref{eq:mdp0}.  For case 2, MDP$_{99}$ is given by
Eq.~\ref{eq:mdp3}.  Thus:

\begin{equation}
{\rm MDP}_{99} =  \frac{4.292 [\int A(E) E^{-\Gamma} dE]^{1/2} }{ [ N \int \mu(E)^X A(E) E^{-\Gamma} dE]^{1/2} } ~~~~.
\label{eq:mdp99}
\end{equation}

\noindent
where $X$ = 1 for case 1 and 2 for case 2 and the integral in the numerator normalizes the
spectrum so that there are $N$ expected counts.
For this exercise, $N = 10^5$ after the ellipticity exclusion,
achievable with \ixpe\ for
sources with 2-8 keV fluxes of $\sim 10^{-11}$ erg cm$^{-2}$ s$^{-1}$ in
about 10 days \citep{ixpe}.
For Case 1, MDP$_{99}$ came out to 3.61\% and 4.52\%
for $\Gamma = 0$ or $2$, respectively.  For the optimum weighting of Case 2, MDP$_{99}$ values were
3.41\% and 4.21\%.  Thus, the improvement to the MDPs were 6.0\% and 7.2\% for the two
different spectral slopes. It is important to note that the fractional improvement is independent of $N$.

One can obtain further improvement in the MDP by increasing the modulation factor.
\citet{PEIRSON2021164740} show that an approach to measuring \ixpe\ tracks using
convolutional neural networks can increase $\mu(E)$, as shown in Fig.~\ref{fig:functions},
and thereby improve the subsequent
estimate of the MDP.
The actual improvement in MDP in this approach depends somewhat on a weighting parameter, $\lambda$
\citep[cf.][]{PEIRSON2021164740}.
Using Eq.~\ref{eq:mdp99} with the $\lambda = 2$ modulation
factor curve
from \citet{PEIRSON2021164740}  and
assuming $N = 10^5$ can yield further reductions of
MDP$_{99}$ by 15\% and 8\% for the two spectral shapes to 2.97\% and 3.89\%, respectively.
For $\lambda = 1$, the improvement is less than a few percent.
\citet{PEIRSON2021164740} point out that the track uncertainty
weight method has no ellipticity exclusion, thus starting with more
events than with the moments method,
but does reduce the effective number of events by 15-20\%.

Clearly, there is an advantage to optimal weighting using the count-weighted
RMS of the modulation factor.  The actual improvement will depend on the details
of the shape of the effective area and modulation factor curves as well as the actual shape
of the spectrum.  However, this weighting can be generally applied for the best results.

\acknowledgments
Funding for this work was provided in part by contract 80MSFC17C0012 from the
Marshall Space Flight Center (MSFC) to MIT in support of IXPE, a NASA Astrophysics Small Explorers mission.
Support for this work was also provided in part by the National Aeronautics and
Space Administration (NASA) through the Smithsonian Astrophysical Observatory (SAO)
contract SV3-73016 to MIT for support of the Chandra X-Ray Center (CXC),
which is operated by SAO for and on behalf of NASA under contract NAS8-03060.
The support of the IXPE instrument Italian team, composed of
INAF and INFN personnel, is gratefully acknowledged.
I thank the anonymous referee for insightful comments that resulted in an improved manuscript.

\bibliographystyle{aasjournal} 

\end{document}